\documentclass[pdflatex,sn-mathphys-num]{sn-jnl}% Math and Physical Sciences Numbered Reference Style
\usepackage{graphicx}%
\usepackage{multirow}%
\usepackage{amsmath,amssymb,amsfonts}%
\usepackage{amsthm}%
\usepackage{mathrsfs}%
\usepackage[title]{appendix}%
\usepackage{xcolor}%
\usepackage{textcomp}%
\usepackage{manyfoot}%
\usepackage{booktabs}%
\usepackage{algorithm}%
\usepackage{algorithmicx}%
\usepackage{algpseudocode}%
\usepackage{listings}%
\usepackage{commath}
\theoremstyle{thmstyleone}%
\theoremstyle{thmstyletwo}%

\theoremstyle{thmstylethree}%

\begin{document}

\title[Article Title]{Large Scale Community-Aware Network Generation}


\author*[1]{\fnm{Vikram} \sur{Ramavarapu}}\email{vikramr2@illinois.edu}

\author[2]{\fnm{João} \sur{Alfredo Cardoso Lamy}}\email{joaoacl@al.insper.edu.br}
\author[3]{\fnm{Mohammad} \sur{Dindoost}}

\author[3]{\fnm{David A.} \sur{Bader}}

\affil*[1]{\orgdiv{Computer Science}, \orgname{University of Illinois at Urbana-Champaign}, \orgaddress{\city{Urbana}, \postcode{61801}, \state{IL}, \country{USA}}}

\affil[2]{\orgdiv{Computer Science}, \orgname{Insper Institute}, \orgaddress{\city{São Paulo}, \state{SP}, \country{Brazil}}}

\affil[3]{\orgdiv{Data Science}, \orgname{New Jersey Institute of Technology}, \orgaddress{\city{Newark}, \postcode{07102}, \state{NJ}, \country{USA}}}


\abstract{Community detection, or network clustering, is used to identify latent community structure in networks. Due to the scarcity of labeled ground truth in real-world networks, evaluating these algorithms poses significant challenges. To address this, researchers use synthetic network generators that produce networks with ground-truth community labels. RECCS is one such algorithm that takes a network and its clustering as input and generates a synthetic network through a modular pipeline. Each generated ground truth cluster preserves key characteristics of the corresponding input cluster, including connectivity, minimum degree, and degree sequence distribution. The output consists of a synthetically generated network, and  disjoint ground truth cluster labels for all nodes. In this paper, we present two enhanced versions: RECCS+ and RECCS++. RECCS+ maintains algorithmic fidelity to the original RECCS while introducing parallelization through an orchestrator that coordinates algorithmic components across multiple processes and employs multithreading. RECCS++ builds upon this foundation with additional algorithmic optimizations to achieve further speedup. Our experimental results demonstrate that RECCS+ and RECCS++ achieve speedups of up to 49× and 139× respectively on our benchmark datasets, with RECCS++'s additional performance gains involving a modest accuracy tradeoff. With this newfound performance, RECCS++ can now scale to networks with over 100 million nodes and nearly 2 billion edges.}

%% \keywords{keyword1, Keyword2, Keyword3, Keyword4}



\maketitle

\section{Introduction}\label{sec1}

Community detection, or network clustering, is the process of partitioning a network into modular, meso-scale subgraphs, and has broad applications across many domains \cite{Fortunato2022-qa, 8625349, Waltman2012-ni}. Community detection algorithms typically operate with respect to an optimality metric that defines a ``good" community. For instance, algorithms like Louvain \cite{Blondel_2008} optimize modularity \cite{Newman2006-fh}, which favors higher edge density within clusters and lower edge density between clusters. The Leiden algorithm \cite{Traag2019-ab} can optimize both Modularity the Constant Potts Model score \cite{Traag2011-le}, which emphasizes more clique-like cluster structure. Connectivity, or minimum cut value (mincut), measures the robustness of a community—the minimum number of edges whose removal would disconnect the cluster. Methods such as the Connectivity Modifier \cite{Park2024-ju} directly optimize this metric. Probabilistic approaches like Stochastic Block Models (SBM) \cite{Peixoto_2019} detect communities through likelihood maximization over block assignments and inter-block edge probability matrices.

\begin{figure}
    \centering
    \includegraphics[width=\linewidth]{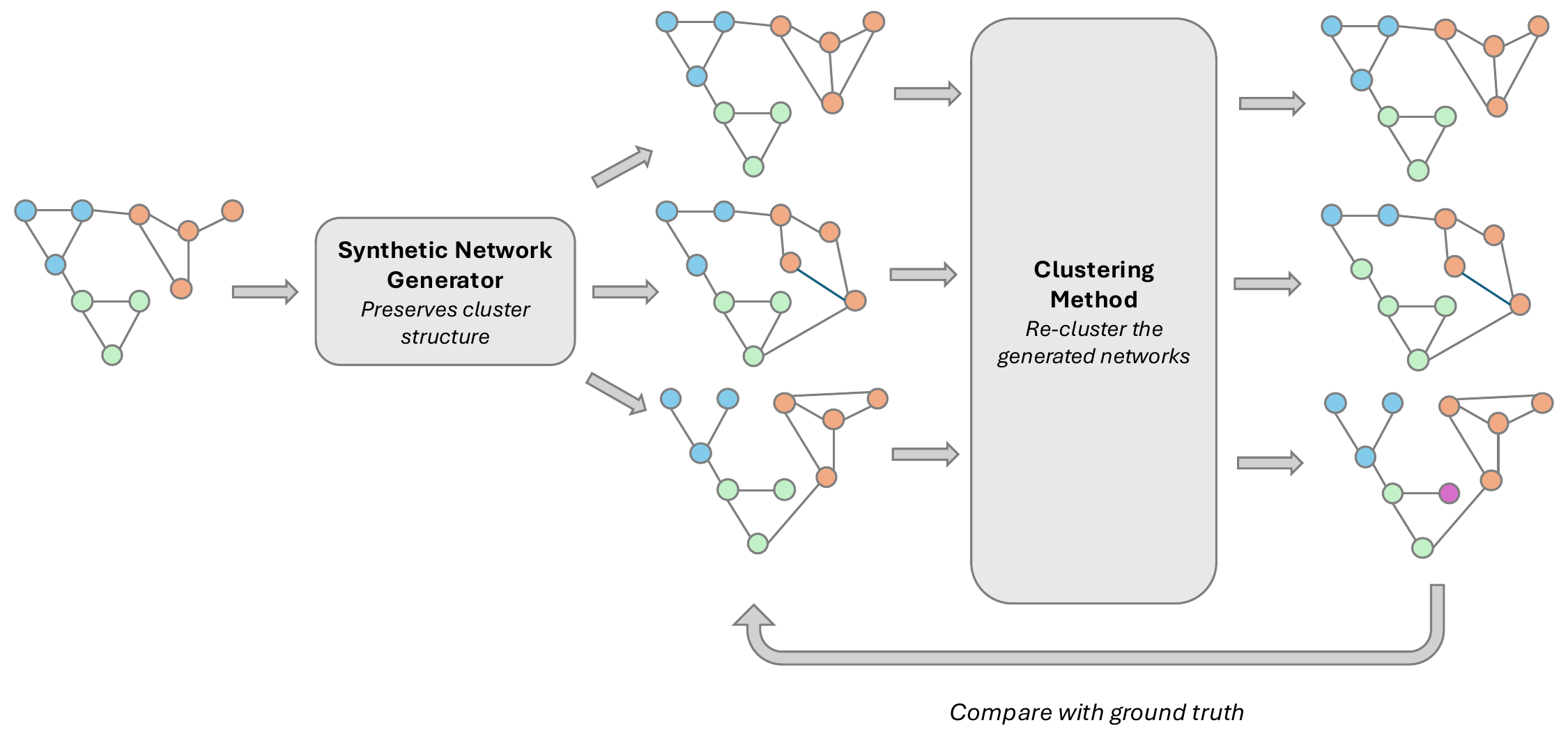}
    \caption{\textbf{A general network clustering evaluation pipeline.} A reference network and clustering is taken in as input to a synthetic network generator, which generates multiple synthetic networks with ground truth communities. When a clustering method is run on these generated networks, they can be compared with the ground truth community labels for accuracy.}
    \label{fig:syn-gen-generic}
\end{figure}

Evaluating community detection algorithms poses significant challenges due to the scarcity of ground truth labels in real-world networks. To address this, researchers use synthetic network generators for benchmarking. When a network and its clustering are provided as input to a synthetic network generator, it produces multiple similar networks with ground truth community labels. These generated networks allow clustering methods to be evaluated by comparing their predictions against the known ground truth. A general evaluation pipeline utilizing a synthetic network generator is illustrated in Figure \ref{fig:syn-gen-generic}. The Lancichinetti-Fortunato-Radicchi (LFR) \cite{Lancichinetti_2008} benchmark generates networks by drawing node degrees and community sizes from power-law distributions, then wiring edges according to a mixing parameter $\mu$ that controls the fraction of each node's edges connecting outside its community. While widely used, LFR generates networks from distributional parameters rather than replicating the specific structural properties of a reference network. The Stochastic Block Model (SBM) offers an alternative approach to synthetic network generation, taking as input block assignments and inter-block edge probability matrices—parameters that are otherwise inferred during community detection—to generate networks with community structure. A degree-corrected variant \cite{PhysRevE.83.016107} additionally incorporates a degree sequence to better model real-world networks. However, SBM-generated networks often contain disconnected or weakly connected clusters, and SBM does not explicitly model important structural properties such as connectivity and minimum degree \cite{anne2024syntheticnetworkspreserveedge}. Edge-Connected SBM (EC-SBM) \cite{The-Anh_Vu-Le2025-cr} addresses this by first generating k-edge-connected subnetworks for each cluster to ensure desired connectivity, then using SBM to generate remaining edges between clusters. For the purposes of this paper, we focus on \textbf{RECCS} \cite{anne2025reccsrealisticclusterconnectivity}, which takes as input a network and its clustering and generates a synthetic network that preserves key characteristics of the input clustering, including connectivity (mincut), minimum degree, and degree sequence distribution.

The RECCS pipeline works as follows: the reference network is divided into clustered and singleton subnetworks. A clustered node is a node that belongs to a cluster of more than one node. The clustered subnetwork, $G_c$ is the induced subgraph of all clustered nodes. The singleton subnetwork $G_s$ is the edge-based complement of $G_c$. In other words, the edge sets of $G_c$ and $G_s$ are disjoint and union to the edge set of the entire network.

Then, a degree-corrected SBM is run on the clustered subnetwork and the singleton subnetwork. The original cluster assignments, sizes, and probabilities are used for the SBM on the clustered subnetwork. For the singleton subnetwork, singletons are treated as their own cluster. Since the singleton subnetwork is an edge set complement of the clustered subnetwork, as opposed to the node set complement, there are still clustered nodes in the singleton subnetwork. As such, clustered nodes in $G_s$ are assigned to their original clusters, while singletons are assigned to their own clusters. This cluster assignment is used to generate a SBM of the singleton subnetwork.

The SBM-generated clustered subnetwork is then processed by the \textbf{RECCS module}, which consists of four stages. In the first stage, edges are added to ensure that each node satisfies a minimum degree requirement per cluster, determined by the minimum cut value of the corresponding reference cluster. Then, in the second stage, edges are added to disconnected clusters to stitch connected components together. In the third stage, edges are added between minimum cut partitions so that the cluster meets the minimum cut requirement. In stage four, edges are added to match degree sequences in the generated clusters to the reference clusters.

In the original implementation of RECCS, stage four is done with two different strategies. The first version uses a max heap prioritizing nodes with the highest degree deficit. The second version computes a degree-corrected SBM using leftover degrees that are appended to the cluster. For the scope of this paper, we focus on improving the performance of the first version of RECCS, and leave the second version for future work.

Once the RECCS module has processed the clustered subnetwork, it is merged with the singleton subnetwork to produce a final graph. The merge is a simple edge set concatenation.

While RECCS is effective in generating synthetic networks that preserve key properties of the input clustering, it suffers from long runtimes, especially on large networks with many clusters. This limits its applicability for large-scale network generation tasks. To address this limitation, we first present \textbf{RECCS+}, which introduces parallelization strategies to the original RECCS pipeline while maintaining algorithmic fidelity. In order to achieve further speedups, we then find that an algorithmic redesign is necessary, leading to the development of \textbf{RECCS++}. RECCS++ incorporates additional algorithmic optimizations alongside parallelization to achieve significant performance improvements, albeit with a modest tradeoff in accuracy.

% The original RECCS end-to-end pipeline, including the RECCS module, which processes the clustered subnetwork, is shown in Figure \ref{fig:reccs-module}.

%In the following sections, we first describe the data used for evaluation in Section \ref{sec2}. Then, we describe our methods in Section \ref{sec3}, including the parallelization strategies used in RECCS+ and the algorithmic optimizations in RECCS++. We present our results in Section \ref{sec12}, including speedup results for RECCS+ and RECCS++ as well as accuracy comparisons to demostrate algorithmic fidelity. In Section \ref{sec14}, we present further algorithmic analysis, demonstrating RECCS, RECCS+, and RECCS++'s use and behavior as a synthetic network generation software by running reclustering evaluation on our generated networks. Finally, we conclude with a summary of our findings and future directions in Section \ref{sec13}.
In the following sections, we first describe the data used for evaluation in Section \ref{sec2}. Then, we describe our methods in Section \ref{sec3}, including the parallelization strategies used in RECCS+ and the algorithmic optimizations in RECCS++. We present our results in Section \ref{sec4}, including speedup results for RECCS+ and RECCS++ as well as accuracy comparisons to demonstrate algorithmic fidelity. In other words, we show that RECCS+ produce networks with near identical properties to those generated by the original RECCS. Since RECCS++ introduces algorithmic changes, we also analyze the differences in network properties between RECCS++ and the original RECCS. In Section \ref{sec5}, we present further algorithmic analysis, demonstrating RECCS, RECCS+, and RECCS++'s use and behavior as a synthetic network generation software by running reclustering evaluation on our generated networks. In Section \ref{sec6}, we present a abstract generalization of RECCS as an algorithmic framework that opens up possibilities for future work. Finally, we conclude with a summary of our findings and future directions in Section \ref{sec7}.

% \begin{figure}
%     \centering
%     \includegraphics[width=\linewidth]{reccs_module_pipeline_orig_full.pdf}
%     \caption{In its original form, the RECCS pipeline starts with parameters derived from an input network and clustering. Then clustered SBM, RECCS module, unclustered SBM, and merge operations are run in serial succession. The RECCS module consists of four stages. Each stage adds edges to: meet minimum degree requirements per node, stitch connected components in disconnected clusters, strengthen connectedness between minimum cut partitions to meet minimum cut requirements, and then match degree sequences, respectively.}
%     \label{fig:reccs-module}
% \end{figure}

\section{Data}\label{sec2}

In our preliminary analysis, we also use the Cit-HepPh dataset from the SNAP dataset \cite{10.1145/2898361}, with 34,546 nodes and 421,578 edges. We also use the Curated Exosome Network (CEN), which is a citation network of publications in exosome biology \cite{10.1162/qss_a_00184}. In the preliminary analysis, we cluster both networks using Leiden with resolution 0.01.

\begin{figure}
    \centering
    \includegraphics[width=\linewidth]{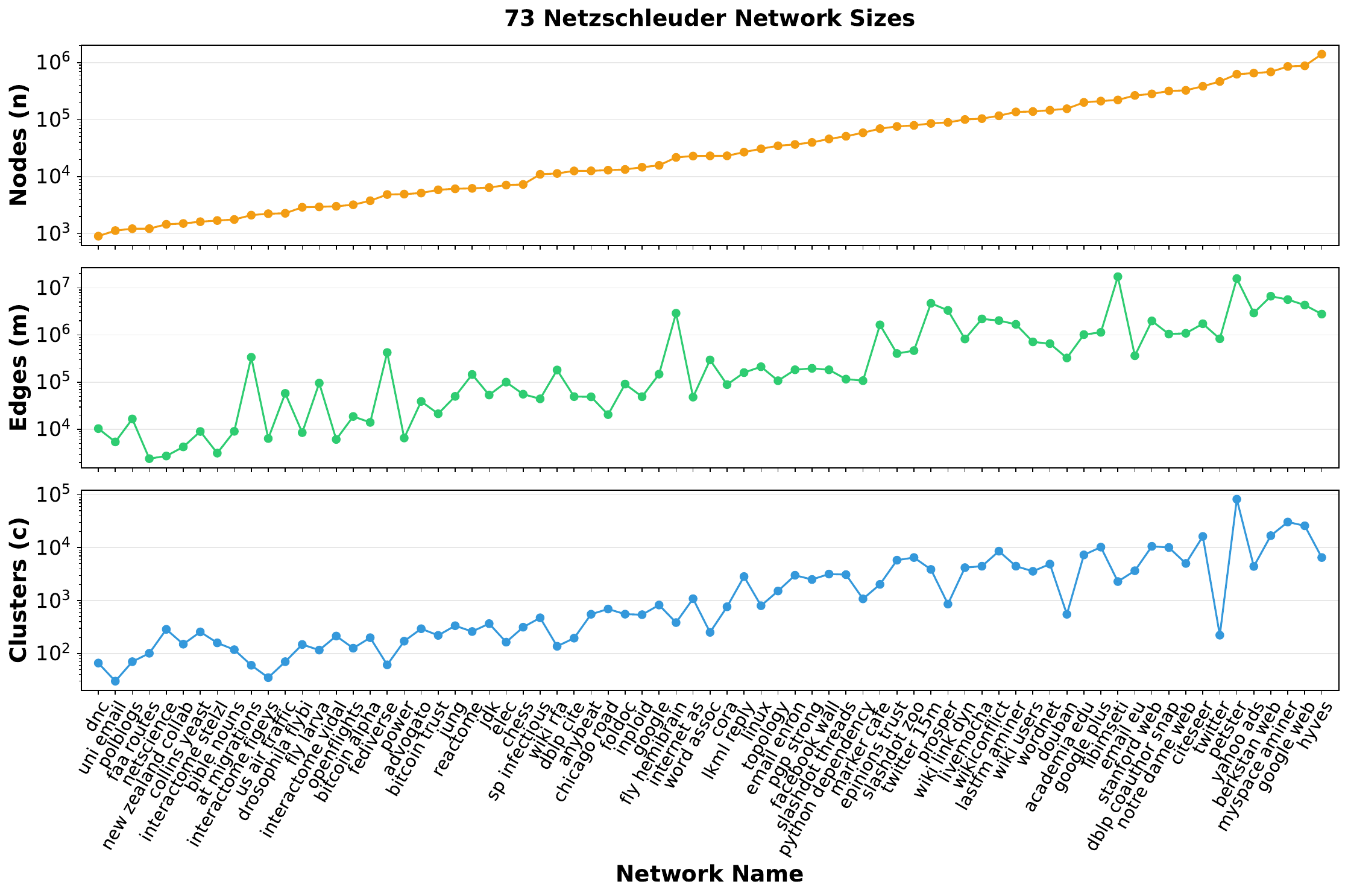}
    \caption{\textbf{Network sizes and number of clusters in the 73 Netzschleuder set.} The top and middle panels show the number of nodes ($n$) and edges ($m$) respectively, while the bottom panel shows the number of clusters ($c$) in each network.}
    \label{fig:network-sizes.}
\end{figure}

For algorithmic fidelity evaluation, we use 73 of the 74 networks from the EC-SBM paper's network list, all of which are from the Netzschleuder network catalogue \cite{Peixoto2020-rt}, as our benchmark datasets for algorithmic fidelity and an initial speedup evaluation. Each network is clustered using SBM-WCC \cite{vule2025usingstochasticblockmodels}. We exclude the \texttt{marvel\_universe} network because our clustering contains only outlier nodes, meaning RECCS would effectively reduce to a degree-corrected Erdos-Renyi generator \cite{Erdos2022-ur}. These networks span a variety of domains, including social, biological, and information networks, ranging from 906 to 1,402,673 nodes and 2,408 to 17,233,144 edges. Each network size is shown in Figure \ref{fig:network-sizes.}.

\begin{figure}
    \centering
    \includegraphics[width=\linewidth]{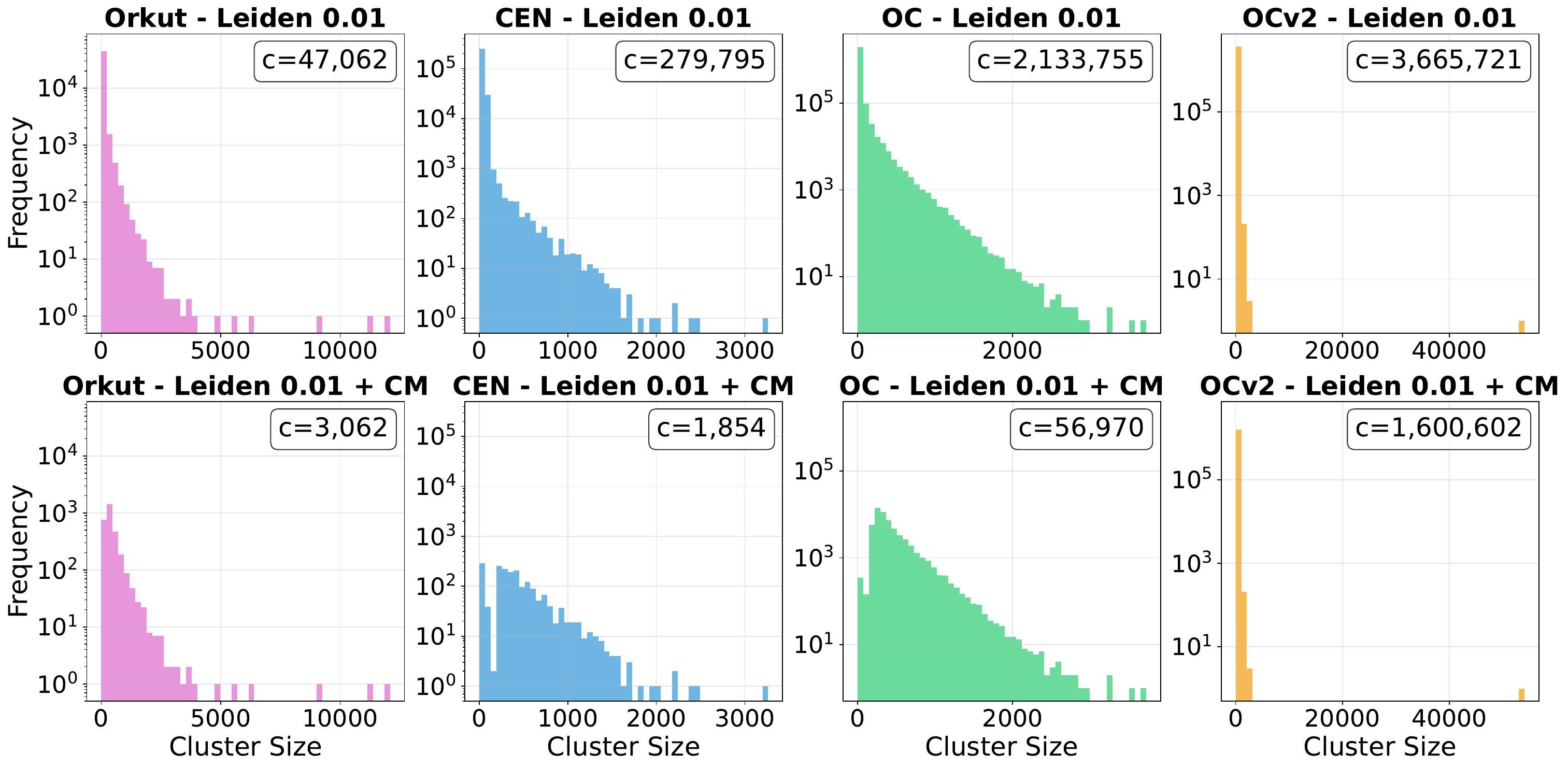}
    \caption{\textbf{Cluster size distribution across all large networks used for scalability experiments.} We use the Orkut, CEN, OC, and OCv2 networks for scalability testing. The networks are clustered with Leiden 0.01, both with and with out Connectivity Modifier (CM) treatment.}
    \label{fig:size_dists}
\end{figure}

We also use four larger networks to evaluate the scalability of RECCS+ and RECCS++: Orkut \cite{10.1145/2898361, yang2012definingevaluatingnetworkcommunities}, the Curated Exosome Network (CEN) mentioned earlier, Open Citations (OC) \cite{illinoisdatabankIDB-6389862, 10.1162/qss_a_00023}, and Open Citations V2 (OCv2) \cite{illinoisdatabankIDB-5216575, 10.1162/qss_a_00023}. Each of these networks were clustered using Leiden with resolution 0.01, both with and without Connectivity Modifier (CM) treatment. We ran CM using our CM++ pipeline software \cite{Ramavarapu2024-wr}. The cluster size distribution across the large networks are shown in Figure \ref{fig:size_dists}.

Finally, we use seven networks from the "Testing 2" dataset in the original RECCS paper for further reclustering accuracy analysis. This includes the following networks: gemsec-Facebook (artist-edges), com-LiveJournal (com-lj), com-Youtube, gemsec-Deezer (HR-edges), twitch-gamers, musae-github and ego-Twitter. All of these networks are from the SNAP catalogue.

All networks used in this study are listed in Table~\ref{tab:large_networks}.

\begin{table}[htbp]
\centering
\caption{\textbf{Network datasets used in this study.} Network sizes are given in number of nodes and edges. The experiments column indicates which experiments each network was used for: Preliminary analysis, Algorithmic Fidelity, Scalability, and Reclustering.}
\label{tab:large_networks}
\begin{tabular}{lrrrr}
\toprule
\textbf{Network(s)} & \textbf{Nodes} & \textbf{Edges} & \textbf{Experiment(s)} & \textbf{Citation(s)} \\
\midrule
Cit-HepPh & 34,546 & 421,578 & Preliminary & \cite{10.1145/2898361, 10.1145/1081870.1081893, 10.1145/980972.980992} \\
Netzschleuder Set & 906–1,402,673 & 2,408–17,233,144 & Fidelity & \cite{Peixoto2020-rt, The-Anh_Vu-Le2025-cr} \\
Orkut & 3,072,441 & 117,185,083 & Scalability & \cite{10.1145/2898361, yang2012definingevaluatingnetworkcommunities} \\
CEN & 13,989,436 & 92,051,051 & Preliminary, Scalability & \cite{10.1162/qss_a_00184} \\
OC & 75,025,194 & 1,363,303,678 & Preliminary, Scalability & \cite{illinoisdatabankIDB-6389862, 10.1162/qss_a_00023} \\
OCv2 & 107,054,145 & 1,962,840,983 & Scalability & \cite{illinoisdatabankIDB-5216575, 10.1162/qss_a_00023} \\
artist-edges & 50,515 & 819,306 & Reclustering & \cite{10.1145/2898361, rozemberczki2019gemsec}\\
% com-Amazon &&& Reclustering & \\
com-lj & 3,997,962 & 34,681,189 & Reclustering & \cite{10.1145/2898361, yang2012definingevaluatingnetworkcommunities} \\
com-Youtube & 1,134,890 & 2,987,624 & Reclustering & \cite{10.1145/2898361, yang2012definingevaluatingnetworkcommunities}\\
HR-edges & 54,573 & 498,202 & Reclustering & \cite{10.1145/2898361, rozemberczki2019gemsec}\\ 
twitch-gamers & 168,114 & 6,797,557
 & Reclustering & \cite{10.1145/2898361, rozemberczki2021twitch} \\
musae-github & 37,700 & 289,003
 & Reclustering & \cite{10.1145/2898361, rozemberczki2019multiscale} \\
% soc-Pokec &&& Reclustering & \\
ego-Twitter & 81,306 & 1,768,149 & Reclustering & \cite{10.1145/2898361, NIPS2012_7a614fd0} \\
\bottomrule
\end{tabular}
\end{table}

\section{Materials and Methods}\label{sec3}

\begin{figure}
    \centering
    \includegraphics[width=0.8\linewidth]{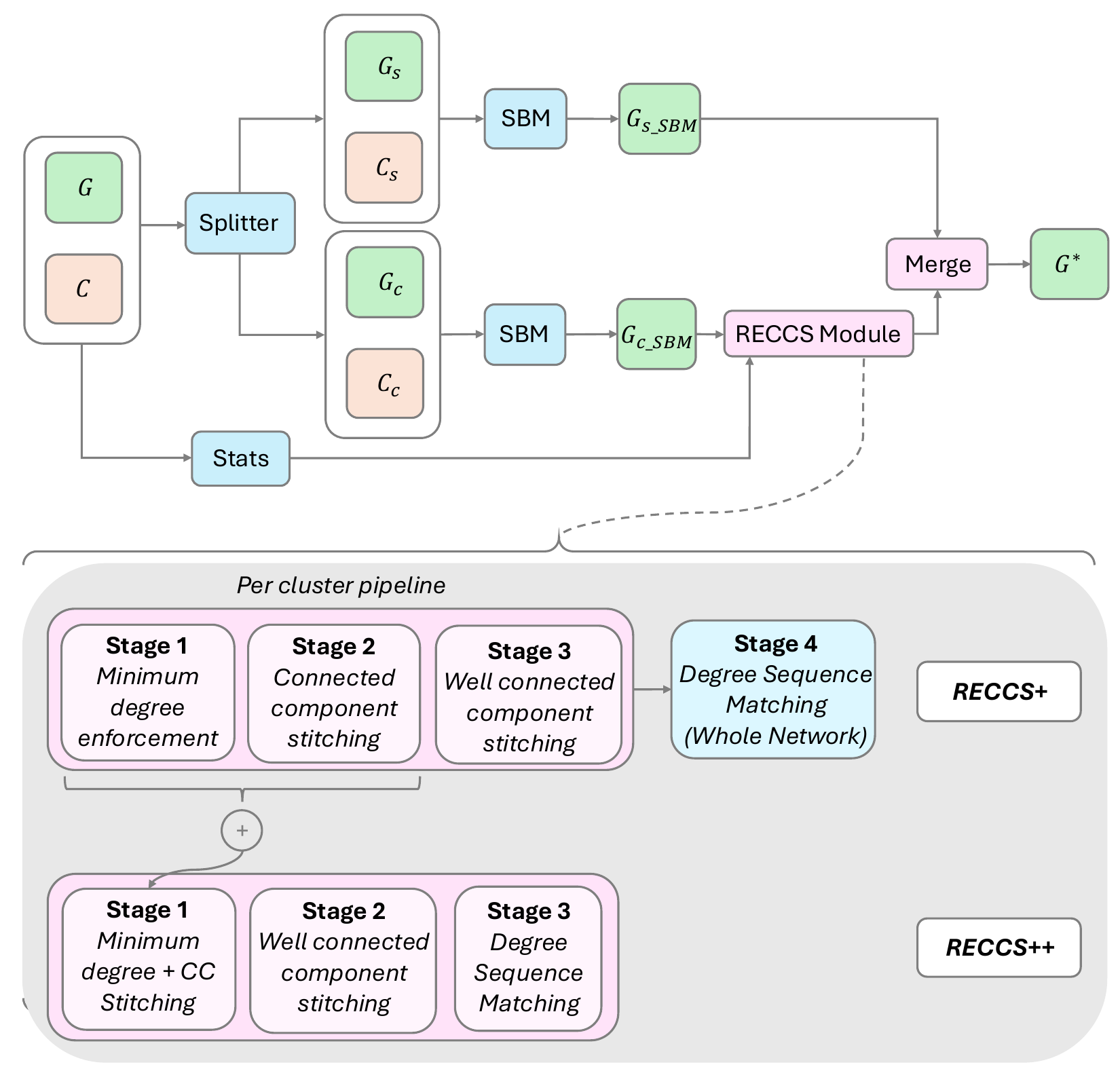}
    \caption{\textbf{The full pipeline of RECCS+ and RECCS++.} First, statistics are collected for the input clustering. In the meantime, a splitter outputs the clustered and singleton sunetworks. When the splitter is finished, SBMs are run on both subnetworks in parallel. When the clustered SBM and the stats are both done, the RECCS module is run. When the RECCS module and the singleton SBM are finished, they are merged into a final output graph. All C++ processes are labeled in pink, while all Python processes are blue.}
    \label{fig:full-pipeline}
\end{figure}

\subsection{Algorithmic Details}\label{sec3.1}
Both RECCS+ and RECCS++ employ a multi-process orchestrator that executes tasks in parallel following the pipeline shown in Figure \ref{fig:full-pipeline}. The pipeline takes as input a reference network and its clustering, replicating both the network topology and cluster statistics in the generated output.

The pipeline begins by splitting the input into two induced subgraphs: $G_c$ contains all nodes belonging to clusters of size greater than one, while $G_s$ contains its edge complement. Both subgraphs inherit their cluster assignments from the input clustering $C$, yielding $C_c$ and $C_s$ respectively. Concurrently, extracts cluster-level statistics (number of nodes and edges and mincut size) from the input to guide the RECCS module. Once the splitting completes, SBMs are run on both $(G_s,C_s)$ and $(G_c,C_c)$ in parallel, generating $G_{s\_SBM}$ and $G_{c\_SBM}$. The clustered SBM, $G_{c\_SBM}$ is processed by the RECCS module, after which the processed result is concatenated with $G_{s\_SBM}$ to produce the final synthetic network $G^*$.

The RECCS module behavior differs between variants. In RECCS+, the module is algorithmically consistent with the original, with parallelism implemented. Minimum degree enforcement, connected component stitching, and well-connected component stitching operate per cluster via a multithreaded task queue, followed by global degree sequence matching after all cluster-level tasks complete. In RECCS++, all operations—including degree sequence matching—are performed per cluster within the task queue across three stages: (1) combined minimum degree enforcement and connected component stitching, (2) well-connected component stitching, and (3) per-cluster degree sequence matching. This per-cluster approach in RECCS++ enables greater parallelization compared to the global degree sequence matching in RECCS+, but results in slightly different network properties of $G^*$.

Minimum degree enforcement and connected component stitching are combined in RECCS++ by prioritizing edge additions that satisfy both requirements. When a cluster is disconnected, stitching edges between components connect the lowest-degree node from each component, after which minimum degree enforcement proceeds as normal.

\subsection{System Details}

For preliminary and reclustering experiments in Sections \ref{implementation} and \ref{sec14}, we used the University of Illinois Campus Cluster \texttt{zgdrasil} partition, using Intel Xeon Platinum 8568Y+ 2.30 GHz processors (96 cores) and 128 GB of RAM. For the algorithmic fidelity and scalability experiments in Sections \ref{sec12} and \ref{sec13}, we used the NJIT Wulver Cluster, with AMD EPYC 7753 2.45GHz CPUs (64 cores). For algorithmic fidelity experiments, we used the \texttt{general} partition with 512GB of RAM, and for scalability experiments, the \texttt{bigmem} partition with 2TB of RAM.

\subsection{Implementation Details}\label{implementation}
In our implementation of RECCS, we primarily use C++ for the RECCS module. We use the \texttt{graph-tool} Python module \cite{Peixoto2017-dn} to compute SBMs. Since the inputs for the SBMs are two-column edgelists and cluster assignments (node, cluster) of nodes in $G_c$ and $G_s$, splitting the graph does not need any network loading and can be reduced to a simple tabular split that can be performed in Python using Pandas. The stats computation is done in a Python script borrowed from the CM++ Pipeline software, modified to only collect cluster IDs, cluster sizes (number of nodes and edges) and minimum cut values.

RECCS in its original form utilizes \texttt{graph-tool} for SBM generation. It first takes a network and clustering as input and computes an edge count matrix between clusters based on the observed inter-cluster and intra-cluster edge patterns in the input network. Using this matrix along with the cluster assignments and degree sequence from the original network, \texttt{graph-tool} generates a synthetic network via a degree-corrected SBM. The edge count matrix is first computed as a dictionary-of-keys (DOK) sparse matrix format, and is converted to a compressed sparse row (CSR) matrix. RECCS+ and RECCS++ improve SBM performance by using a coordinate (COO) matrix instead of a DOK. While DOK outperforms COO for incremental modifications (changing individual elements one at a time), COO is better for bulk modifications (changing chunks at a time). Therefore, edges are partitioned into chunks to modify the COO-form edge count matrix. Moreover, chunks are handled in parallel via multithreading to further improve matrix construction performance. 

\begin{figure}
    \centering
    \includegraphics[width=0.7\linewidth]{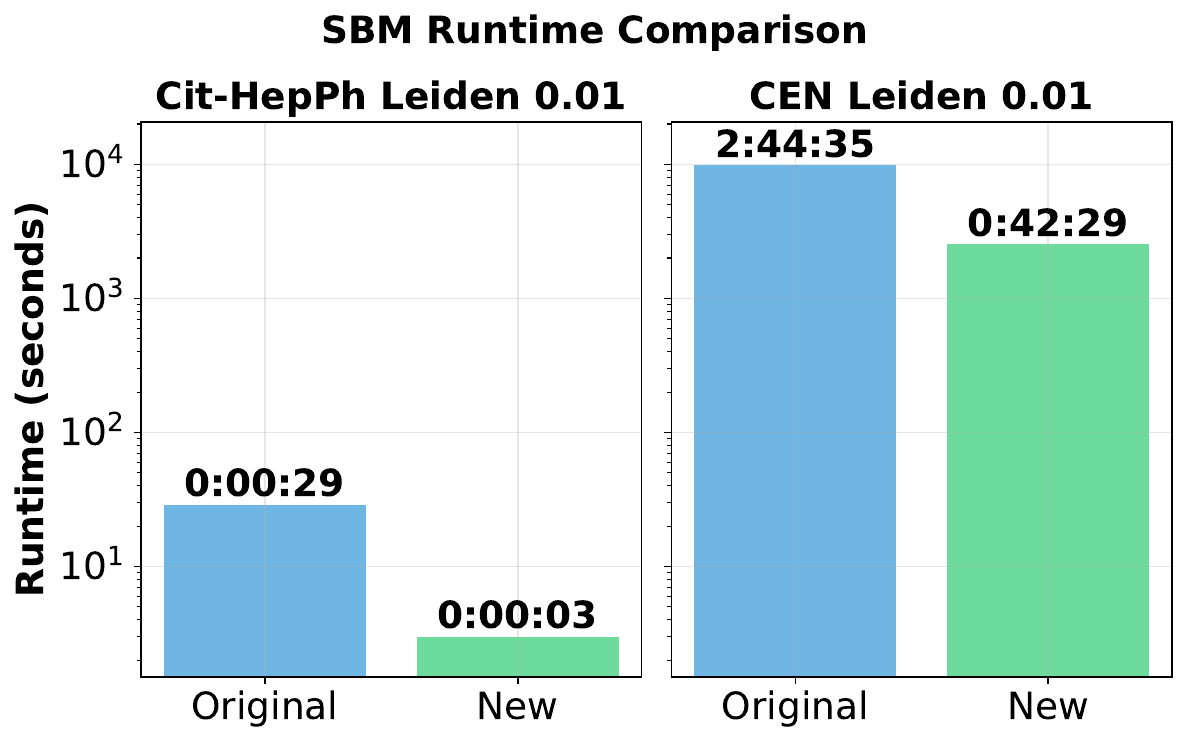}
    \caption{\textbf{Speedup of the SBM stage after optimization.} Algorithmic, data structure, and parallelization optimizations were used to speed up SBM generation. New SBM runs utilize 64 threads. Speedup: Cit-HepPh: 9.7x, CEN: 3.9x.}
    \label{fig:sbm-speedup}
\end{figure}

We verify that these changes made to the SBM generation stage indeed yields significant speed-ups through a preliminary small-scale experiment. For this experiment, we use both Cit-HepPh and CEN, clustered using Leiden CPM with resolution 0.01. As shown in Figure \ref{fig:sbm-speedup}, these changes yield a 9.7x speedup on Cit-HepPh and a 3.9x speedup on CEN.

\begin{figure}
    \centering
    \includegraphics[width=0.5\linewidth]{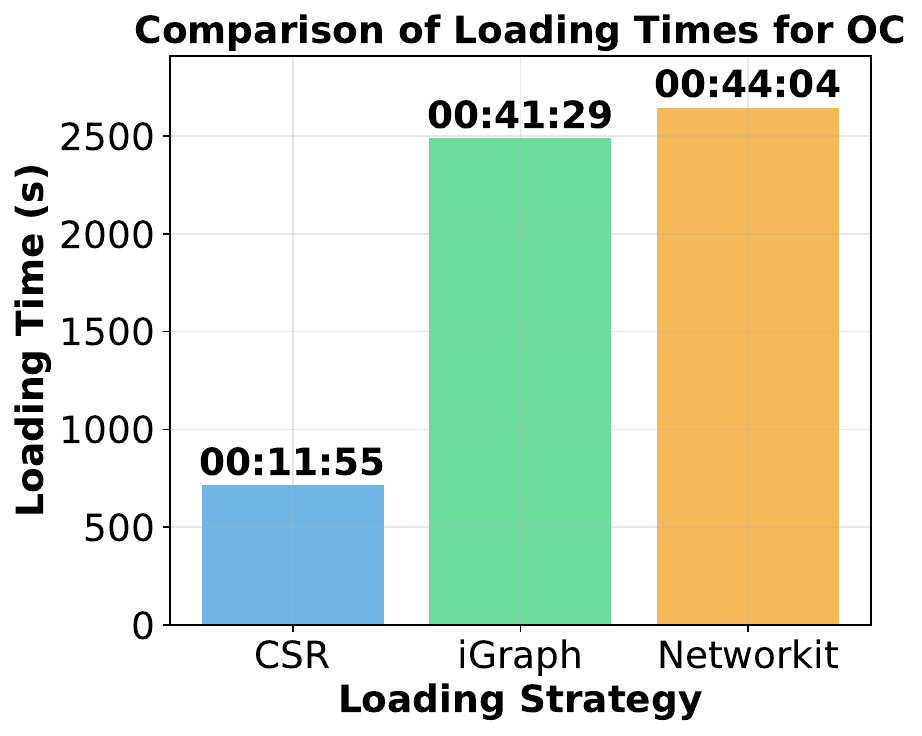}
    \caption{\textbf{CSR graph loader runtime compared to iGraph and Networkit baselines.} All runtimes are single-threaded. CSR Speedup with respect to iGraph: 3.48x, Networkit: 3.70x.}
    \label{fig:oc-loading-time}
\end{figure}

Within the RECCS module, we represent graphs using adjacency matrices in CSR format to improve loading speed and preserve memory using a compressed format. We verify that graph loading is indeed more performant with CSR adjacency representations through a preliminary experiment. We used the OpenCitations network and compared our C++ graph loader runtime with iGraph \cite{igraph, igraph2} and Networkit \cite{staudt2015networkittoolsuitelargescale} baselines. Figure \ref{fig:oc-loading-time} shows that our CSR loader yields a 3.48x performance improvement over iGraph and a 3.70x performance improvement over Networkit.

Moreover, the RECCS module implements shared-memory, single-node parallelism using OpenMP for multithreaded queue operations. With respect to multiprocessing, the process DAG shown in Figure \ref{fig:full-pipeline}—comprising the splitter, stats, and SBM processes—is coordinated through an orchestrator that manages process forks. The splitter process divides the input graph into clustered and unclustered components, and SBMs are run in parallel on both components. During the splitter's runtime, stats are computed on the input network and clustering. The stats script is reused from the CM++ pipeline. In RECCS+, when the multithreaded queue is completed, the degree sequence matching is done through a forked Python process. The degree sequence matching script is reused from the EC-SBM source code\footnote{https://github.com/illinois-or-research-analytics/ec-sbm/blob/main/ec-sbm/correct\_degree.py}.

The code for this software is made available on GitHub\footnote{https://github.com/illinois-or-research-analytics/reccs}

\section{Results}\label{sec4}

\subsection{Algorithmic Fidelity}\label{sec12}

\begin{figure}
    \centering
    \includegraphics[width=\linewidth]{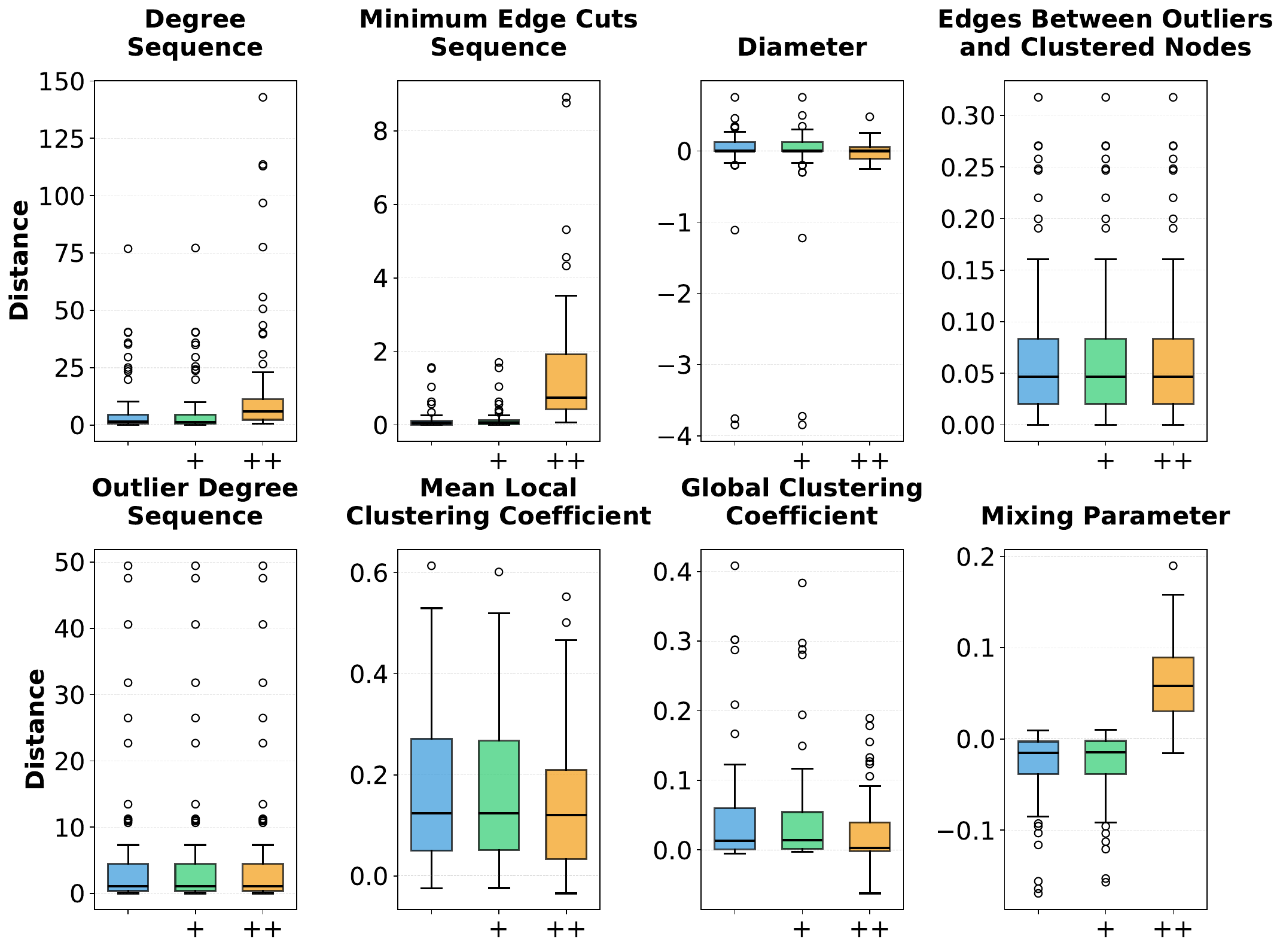}
    \caption{\textbf{Accuracy of network statistics across 73 Netzschleuder networks.} Statistics of the networks generated by RECCS, RECCS+, and RECCS++ (indicated by '', '+', and '++' respectively), with reference to the 73 Netzschleuder networks. The y-axis shows the error metric for each statistic, as described in Table \ref{tab:stats_desc}. Lower values indicate better fidelity.}
    \label{fig:fidelity-comparison}
\end{figure}

A comparison of network statistics across the 73 selected EC-SBM networks are shown in figure \ref{fig:fidelity-comparison}. We use relative difference to evaluate diameter and the number of edges between the clustered and outlier subnetworks. We use absolute difference to evaluate mixing parameter and clustering coefficients. We use root mean-squared error (RMSE) for evaluation of the degree sequence, both across the graph and in the outlier subnetwork, and the minimum edge cut sequence. The absolute and relative differences are used when the real and synthetic network statistics are represented by scalars $s$ and $s'$. RMSE is used when the real and synthetic network statistics are represented by sequences $s$ and $s'$ where the $i^{\text{th}}$ element of both sequences are represented as $s_i$ and $s'_i$. The equations for each metric are represented in equations \ref{absolute_diff}, \ref{rel_diff}, and \ref{rmse}.

\begin{equation}\label{absolute_diff}
    \text{absolute difference} = s - s'
\end{equation}

\begin{equation}\label{rel_diff}
    \text{relative difference} = (s - s')/s
\end{equation}

\begin{equation}\label{rmse}
    \text{RMSE} = \sqrt{\frac{1}{N}\sum_{i=1}^{N}(s_i-s_i')^2}
\end{equation}

\begin{table*}[t]
\centering
\caption{\textbf{Properties of the real and synthetic networks and the metrics used to evaluate them.}}
\label{tab:stats_desc}
\small
\begin{tabular}{lr}
\toprule
\textbf{Stat} & \textbf{Metric Used} \\
\midrule
\textbf{Degree Sequence} & RMSE \\
\textbf{Minimum Edge Cuts Sequence} & RMSE \\
\textbf{Diameter} & Relative \\
\textbf{Edges Between Outliers \& Clustered Nodes} & Relative\\
\textbf{Outlier Degree Sequence} & RMSE \\
\textbf{Mean Local Clustering Coefficients} & Absolute\\
\textbf{Global Clustering Coefficients} & Absolute\\
\textbf{Mixing Parameter} & Absolute \\
\bottomrule
\end{tabular}
\end{table*}

As the purpose of this experiment is to evaluate algorithmic fidelity, all RECCS variants use the same SBMs to avoid differences that may come from randomization. As shown in the figure, there is high algorithmic fidelity between RECCS+ and RECCSv1. However, RECCS++ shows some tradeoff in accuracy. Clusters of RECCS++ exhibit higher connectivity than the original network, more so than the RECCS and RECCS+ networks. Since the degree sequence of RECCS++ exhibit higher RMSE than the other two versions, this is likely due to RECCS++ being constrained to add edges within a cluster to fit degree sequence, therefore leading to higher minimum cut in the clusters of RECCS++ networks than the original. This constraint may also lead to lower mixing parameters in RECCS++ networks as well. While these discrepancies exist, there are some tradeoffs. RECCS++ networks are closer in diameter to the original network than RECCS and RECCS+. RECCS++ also exhibits slightly better accuracy in the clustering coefficients than the other two versions.

\subsection{Performance and Scalability}\label{sec13}

% Figure \ref{fig:runtime-boxplot-comparison} compares the runtimes across the 73 selected EC-SBM networks, with the x-axis showing runtime in seconds on a logarithmic scale. RECCS++ achieves an average 9× speedup over the original RECCS at smaller scales, while RECCS+ achieves an average 4× speedup. Notably, the original RECCS exhibits substantially higher variance in runtime, whereas RECCS+ and RECCS++ demonstrate more consistent performance with tighter runtime distributions and more predictable speedups.

% \begin{figure}
%     \centering
%     \includegraphics[width=\linewidth]{runtime_comparison_boxplot.pdf}
%     \caption{Runtime distributions for RECCS, RECCS+ and RECCS++ across 73 EC-SBM input networks.}
%     \label{fig:runtime-boxplot-comparison}
% \end{figure}

\begin{figure}
    \centering
    \includegraphics[width=0.9\linewidth]{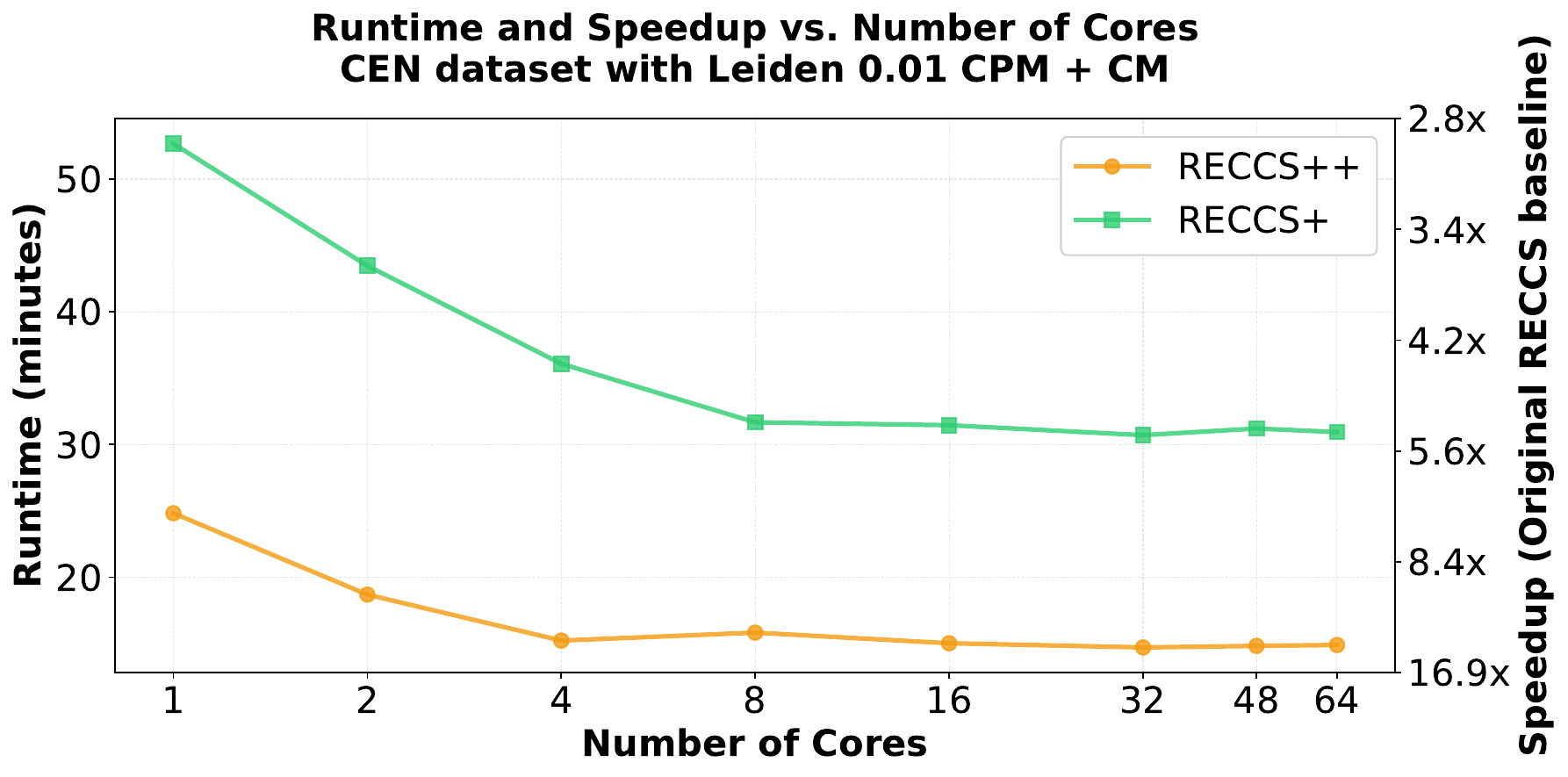}
    \caption{\textbf{End-to-end runtime and speedup of RECCS+ and RECCS++ with respect to RECCS} on CEN, clustered with Leiden 0.01 + CM. Runtime is marked on the left y-axis, while speedup is marked on the right y-axis.}
    \label{fig:cen-speedup-curve}
\end{figure}

Speedup curves of RECCS+ and RECCS++ on the Leiden CPM clustering of CEN with resolution 0.01, treated with CM are shown in Figure \ref{fig:cen-speedup-curve}. The left axis shows runtime in seconds, and the right axis shows speedup relative to the original RECCS runtime on the same network-clustering pair. At 1 core, RECCS+ and RECCS++ achieve 2.9× and 6.8× speedups respectively, increasing to 5.8× and 11.5× at 64 cores. Since parallelism contributes approximately 2–3× speedup from 1 to 64 cores, the majority of the performance gain is attributable to other optimizations, such as the multiprocessed DAG formulation and CSR graph representation.

\begin{figure}[!htbp]
    \centering
    \includegraphics[width=\linewidth]{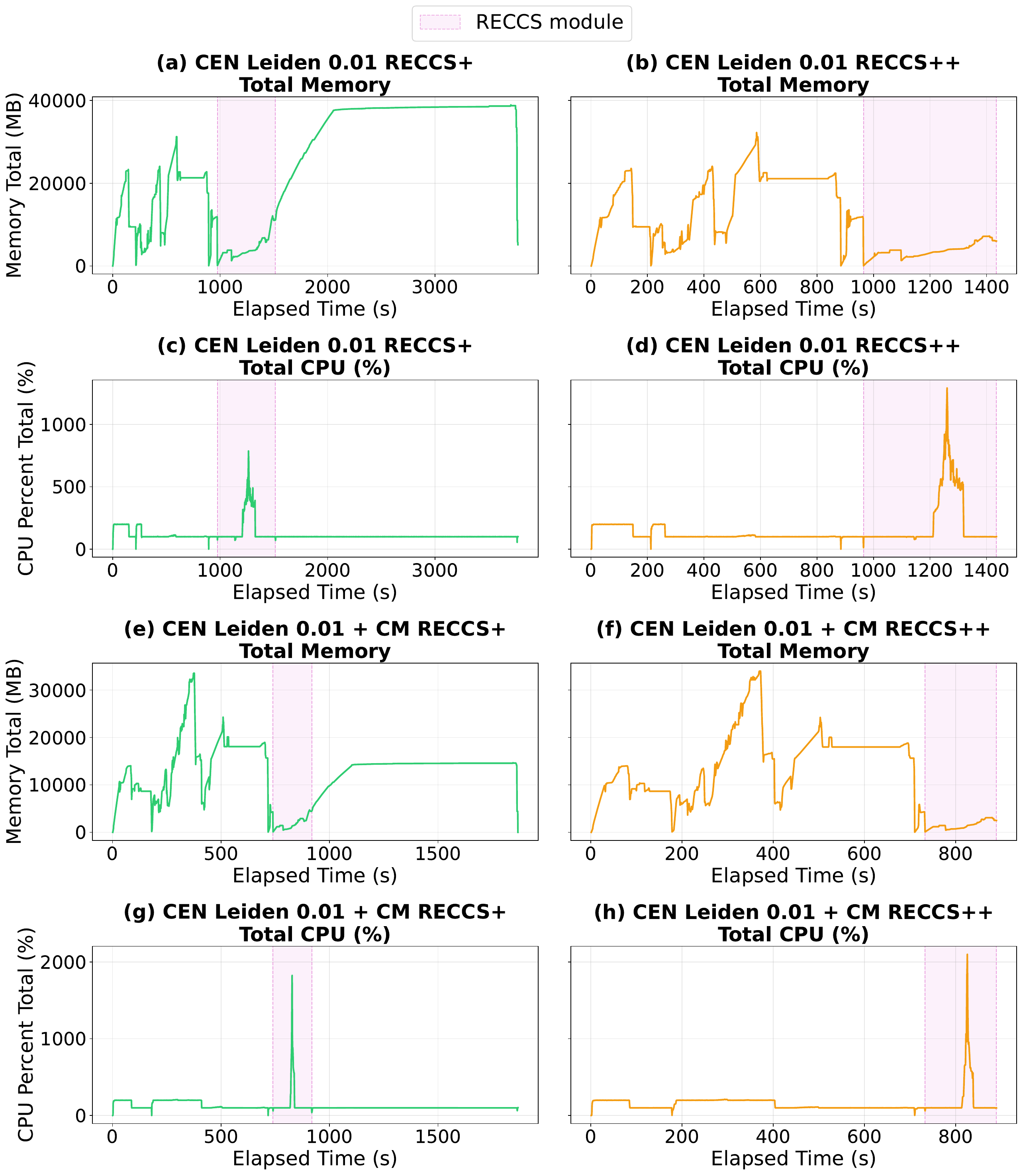}
    \caption{\textbf{Profiling of RAM and CPU usage during the RECCS+ and RECCS++ runtimes} on CEN with a Leiden 0.01 clustering, both with and without CM treatment. The pink shaded region indicates the duration of the RECCS C++ module runtime. CEN Leiden 0.01 RECCS+ (a, c) uses a peak of 38.95GB overall and 12.13 GB during the RECCS module, and a peak of 786.3\% CPU both overall and during the RECCS module. CEN Leiden 0.01 RECCS++ (b, d) uses a peak of 32.28GB overall and 7.2215GB during the RECCS module, and a peak of 1291.3\% CPU both overall and during the RECCS module. CEN Leiden 0.01 + CM RECCS+ (e, g) uses a peak of 33.56GB overall and 4.73GB during the RECCS module, and a peak of 1824\% CPU both overall and during the RECCS module. CEN Leiden 0.01 + CM RECCS++ (f, h) uses a peak of 33.99GB overall and 3.11GB during the RECCS module, and a peak of 2098.7\% CPU both overall and during the RECCS module.
    }
    \label{fig:profiling-comparison}
\end{figure}

To further understand the details of the RECCS+ and RECCS++ runtimes, we used WandB\footnote{https://wandb.ai} to profile the CPU and RAM usage of the two RECCS variants. Figure \ref{fig:profiling-comparison} shows the profiling results of RECCS+ and RECCS++ runtimes on the CEN network with Leiden 0.01 clustering, both with and without CM treatment. The pink shaded region shows the duration of the RECCS C++ module runtime. Both the lowest points in memory usage, and the highest points in CPU usage occur during the actual RECCS module runtime. The lower memory usage trend is likely due to the CSR graph representation taking less memory than the more general purpose \texttt{graph-tool} alternative. Higher CPU usage is due to the highest multithreading engagement being within the RECCS module, more so than during the SBM computation.

Moreover, there is a large portion of RECCS+ runtime after the RECCS module that is not present in RECCS++. This is due to the degree sequence matching stage in RECCS+ being run as a separate Python process after the RECCS module, while in RECCS++ the degree sequence matching is integrated into the RECCS module itself. This stage can take more than half of the RECCS+ runtime, as seen in Figure \ref{fig:profiling-comparison}(a) and \ref{fig:profiling-comparison}(c), and nearly half of the RECCS+ runtime in Figure \ref{fig:profiling-comparison}(e) and \ref{fig:profiling-comparison}(g). Therefore, much of the speedup of RECCS++ over RECCS+ can be attributed to this integration.

Table \ref{tab:benchmark_results} shows the scaling results on the larger networks: Orkut, CEN, OC, and OCv2. RECCS+ and RECCS++ achieve their best performance improvement on CEN clustered with Leiden 0.01, with RECCS+ achieving a 49× speedup over RECCS and RECCS++ achieving a 139× speedup. Notably, on the two largest networks (OC and OCv2), both RECCS and RECCS+ failed to complete within the allocated time window, while RECCS++ successfully completed both networks in under 12 hours. This demonstrates RECCS++'s ability to scale to previously intractable network sizes, processing networks with over 100 million nodes and nearly 2 billion edges.

\begin{table*}[t]
\centering
\caption{\textbf{Table of runtimes per RECCS version across all large network datasets.} Times are in d-hh:mm:ss format. DNC means 'Did Not Complete'.}
\label{tab:benchmark_results}
\small
\begin{tabular}{llccc}
\toprule
\textbf{Network} & \textbf{Clustering} & \textbf{RECCS} & \textbf{RECCS+} & \textbf{RECCS++} \\
\midrule
\textbf{Orkut} & Leiden $r=0.01$ + CM& 6:47:11 & 2:18:08 & 22:44 \\
& Leiden $r=0.01$ & 8:33:41 & 2:29:13 & 23:51 \\
\midrule
\textbf{CEN} & Leiden $r=0.01$ + CM& 2:48:48 & 30:53 & 15:02 \\
& Leiden $r=0.01$ & 2-5:23:59 & 1:04:17 & 23:35 \\
\midrule
\textbf{OC} & Leiden $r=0.01$ + CM& DNC & DNC &  6:08:46 \\
& Leiden $r=0.01$ & DNC & DNC & 7:02:58\\
\midrule
\textbf{OCv2} & Leiden $r=0.01$ + CM& DNC & DNC & 7:59:06 \\
& Leiden $r=0.01$ & DNC & DNC & 11:16:12 \\
\bottomrule
\end{tabular}
\end{table*}

\subsection{Reclustering Evaluation}\label{sec14}
In this experiment, we use synthetic network generation with the exact purpose outlined in the beginning of this paper: to evaluate community detection algorithms with respect to a generated ground truth. As such we follow a reclustering evaluation pipeline like the one shown in Figure \ref{fig:syn-gen-generic}. We take the seven networks from Table \ref{tab:large_networks} designated for the reclustering experiment. These networks are clustered with Leiden CPM with resolution 0.01, Leiden optimizing modularity, and Iterative K-Core clustering (IKC) \cite{10.1162/qss_a_00184}, both with and without CM treatment. RECCS, RECCS+, and RECCS++ networks are generated for each network, clustering and treatment. Then each generated network is reclustered with one of the three methods. If the original clustering used CM treatment, the reclustering will use CM treatment as well. For each clustering reclustering pair, both the normalized mutual information (NMI) and the adjusted rand index (ARI) are computed. 

The NMI and ARI metrics are then averaged across the networks, creating an mean value per clustering-reclustering pair. From this, we derive 12 heatmaps: one per combination of version (3),  treatment (2), and metric (2). Each heatmap is of size 3x3 as each row is a clustering method and each column is a reclustering method. 

Then, we use the Frobenius norm to derive four 3x3 heatmaps to show the difference across RECCS versions across NMI, ARI, NMI with CM treatment, and ARI with CM treatment. The Frobenius norm is a normalized matrix difference, where we take two heatmaps as input:

$$\norm{A-B}_F=\sqrt{\sum_{i=1}^3\sum_{j=1}^3(A_{ij}-B_{ij})^2}$$

\begin{figure}[!htbp]
    \centering
    \includegraphics[width=\linewidth]{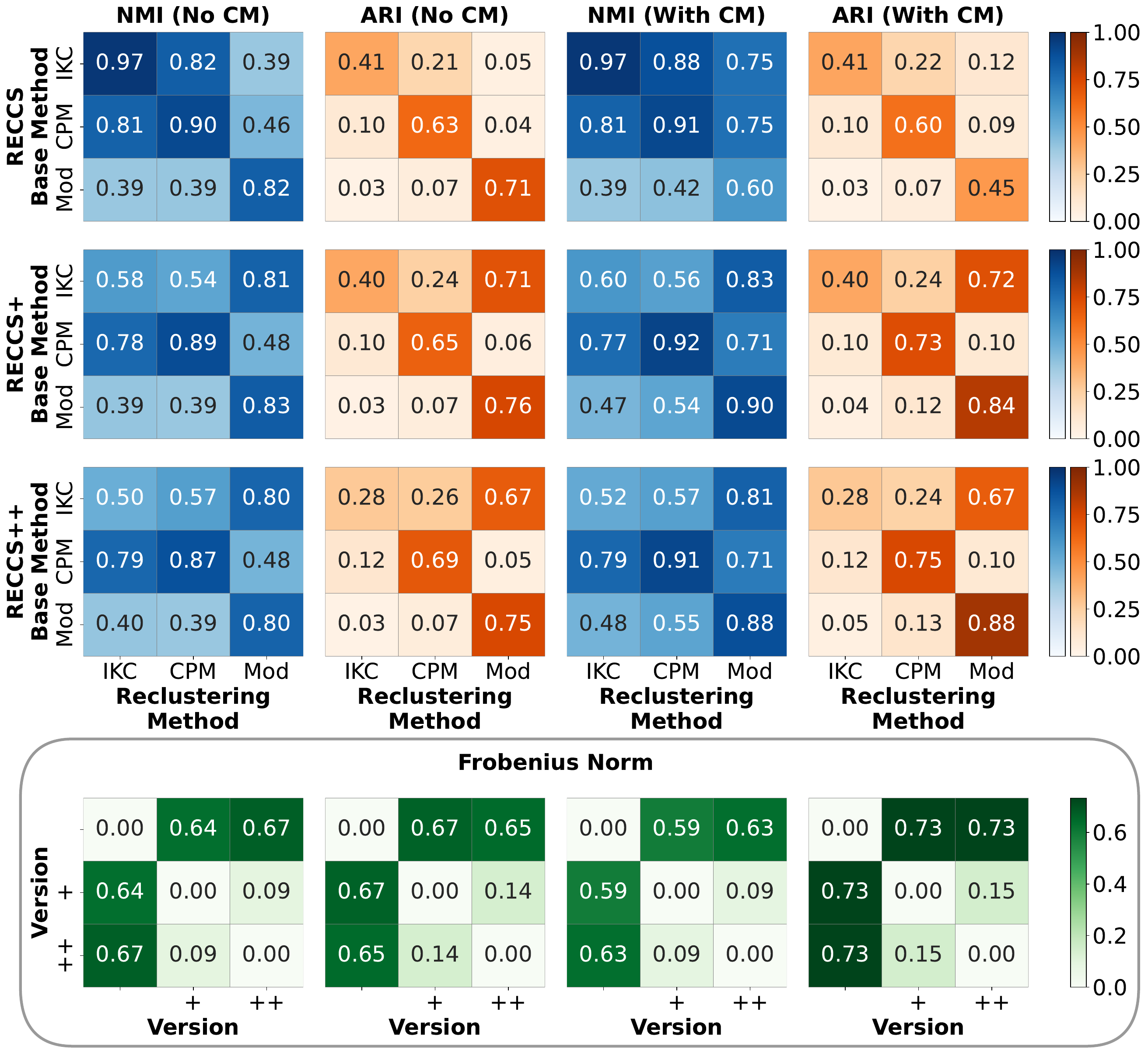}
    \caption{\textbf{Reclustering accuracy comparison across RECCS, RECCS+, and RECCS++}. Rows 1--3 show NMI (blue) and ARI (red) confusion matrices for RECCS, RECCS+, and RECCS++ respectively, comparing original clusterings (IKC, Leiden CPM 0.01, Leiden Modularity) against reclustering methods. Columns 1--2 show results without CM treatment, columns 3--4 with CM treatment. Bottom row (green) shows pairwise Frobenius norm differences between RECCS variants for each metric and treatment condition.}
    \label{fig:reclustering_heatmap}
\end{figure}

Where $A$ and $B$ are two confusion matrices of the same metric across any of the three RECCS versions. 
The final heatmap grid is shown in Figure \ref{fig:reclustering_heatmap}. All heatmaps in blue show a per-version NMI heatmap comparing base clusters to their reclusterings. All heatmaps in red show ARI comparisons. Frobenius norm heatmaps comparing RECCS are shown in green in the bottom row of the grid.

The Frobenius norm matrices show that the confusion matrices of RECCS+ and RECCS++ show more similarity with each other than with the original RECCS. However, the top three rows show that the deviation in matrix similarity mainly comes from the IKC clustering results. This can be due in part to either the generator itself or the clustering method. However, since both Leiden CPM and Leiden Modularity show much smaller Frobenius norm differences across RECCS versions, it is likely that the discrepancy is mainly due to the sensitivity of IKC to network structure changes, as it relies on k-core decomposition which can be significantly affected by even small perturbations in edge structure. In contrast, Leiden CPM and Leiden Modularity appear more robust to the structural variations introduced by different RECCS versions, resulting in more consistent clustering outcomes across synthetic networks.

Despite preserving key network statistics (as shown in Section \ref{sec4}), RECCS+ and RECCS++ produce different reclustering patterns than RECCS, particularly for IKC. This demonstrates that reclustering evaluation can reveal clustering algorithm sensitivities to structural variations beyond standard network metrics. The consistency between RECCS+ and RECCS++ results, combined with their computational efficiency and ability to scale to larger networks, validates their use for this type of robustness analysis in reclustering evaluation, and opens the door for such analysis on larger-scale networks that were previously infeasible with RECCS.

% \clearpage

\section{Discussion: Scalability Enables New Directions}\label{sec5}

The scalability achievements demonstrated in the previous sections, particularly RECCS++'s ability to handle networks with over 100 million nodes, fundamentally expand the potential applications of the RECCS methodology beyond its original evaluation-focused design. While the following discussion applies conceptually to the RECCS approach as a whole, RECCS++'s superior scalability makes it the most suitable implementation for these applications. In this section, we explore how viewing the RECCS methodology through an encoder-decoder lens reveals opportunities for broader impact in network generation tasks.

\subsection{The Encoder-Decoder Abstraction}

Given that the RECCS methodology has achieved scalability through RECCS+ and RECCS++, there are new doors that have been opened with respect to its generalization. The RECCS approach can be conceptualized as an encoder-decoder architecture, analogous to models widely used in image generation \cite{ho2020denoisingdiffusionprobabilisticmodels, Kingma_2019} and increasingly adopted in graph machine learning \cite{kipf2016variationalgraphautoencoders, liu2023generativediffusionmodelsgraphs}. In traditional encoder-decoder models, an encoder embeds training data into a latent space, while a decoder generates new data from these latent embeddings. The RECCS methodology follows a similar paradigm as a simplified, non-deep encoder-decoder framework for network generation. The encoding phase comprises the statistics and cluster probability matrix computations that precede the RECCS module and SBMs, respectively, as these steps extract network and cluster properties that serve as latent features. The RECCS module and SBMs then function as the decoder, generating synthetic networks conditioned on these latent representations.

\subsection{Extensions and Adaptations}

The RECCS methodology has a limitation in that it is a single-reference network generator designed to mimic the original network's characteristics. Without constraints on perturbation, the optimal synthetic network that the RECCS approach could generate would simply be the original reference network itself, which limits its utility as a benchmark for evaluating community detection algorithms on diverse network structures. However, this encoder-decoder abstraction enables several natural extensions of the RECCS methodology beyond its current single-network formulation, and can open up possibilities to perturbations in the network for more robust evaluation. First, the RECCS approach can be adapted to generate scaled versions of input networks through normalization in the encoding phase. Rather than storing absolute network properties, the encoder would extract normalized or spectral properties such as spectral mincut values instead of raw mincut counts, and density metrics instead of explicit node and edge counts. The decoder could then rescale these normalized features by a dilation or shrinkage factor to pass into the RECCS module to  generate networks of varying sizes while preserving structural characteristics.

Second, the RECCS methodology could be extended to learn from distributions of networks rather than individual instances. In this formulation, the encoder would aggregate statistics across multiple training networks to infer distributions over key structural characteristics: cluster size distributions, inter-cluster edge probability matrices, and per-cluster mincut distributions. The decoder would then sample values from these learned distributions and pass them into the RECCS module to generate diverse synthetic networks that capture the variability present in the training data. This extension would address a key limitation of the current approach, which requires a single input network and may therefore impact the robustness and generalizability of synthetic network generation for evaluation purposes. While deep generative models can learn from network distributions, the demonstrated scalability of the RECCS module suggests this approach could offer improved performance over existing deep learning methods, particularly for large network generation tasks.

Lastly, the encoder-decoder abstraction of the RECCS methodology naturally positions RECCS++ as a complementary component to deep learning methods for network generation. While deep learning models excel at learning complex distributions from training data, they face computational challenges when generating large-scale networks directly. The RECCS module and SBM components of RECCS++ address this limitation by providing a scalable decoder that can efficiently generate networks from learned parameters. In this hybrid architecture, the deep learning component would serve as the encoder, learning to infer these distributional parameters from a corpus of training networks. The RECCS module and SBMs from RECCS++ would then serve as the decoder, taking these learned parameters to generate networks efficiently. This division of labor exploits the complementary strengths of both approaches: deep learning's capacity for representation learning and pattern recognition, paired with RECCS++'s demonstrated efficiency and scalability for large-scale generation. Such a hybrid framework could enable generation of synthetic networks at scales currently intractable for purely deep learning-based methods, extending network generation capabilities to networks with hundreds of millions of nodes. Moreover, the deterministic nature of RECCS++ as a decoder provides interpretability advantages over end-to-end deep generative models, as the network generation process remains transparent and the learned parameters directly correspond to structural properties of interest.

\subsection{Challenges and Future Directions}

However, implementing such distribution-based generation presents several technical challenges. The primary difficulty lies in the interdependencies among network properties: cluster sizes, mincut values, and edge densities cannot be sampled independently, as they are structurally constrained by one another. For instance, the dimensionality of the mincut distribution depends on the number of clusters, which itself varies with network size. This creates a hierarchy of coupled distributions—a "distribution of distributions"—where sampling decisions at one level constrain the valid range of values at another. Additionally, when scaling networks through dilation or shrinkage, the number of clusters typically changes non-linearly with network size, requiring variable-length distributions that adapt to the target scale. Addressing these challenges would require careful modeling of the joint distribution over network properties, potentially through hierarchical generative models or constraint-aware sampling procedures. Despite these complications, the encoder-decoder abstraction provides a principled framework for such extensions, making the RECCS methodology readily adaptable to more sophisticated generation scenarios as it matures.

\section{Limitations and Future Work}\label{sec6}

In the previous section, several limitations were touched upon, but they all were inherent to the nature of RECCS, RECCS+ and RECCS++ being single reference network synthetic network generators. As such, we proposed potential future work in developing an abstraction of the RECCS methodology to adapt it to problems that avoid these limitations.

As shown in Figure \ref{fig:profiling-comparison} and discussed in Section \ref{sec13}, memory profiling reveals that the RECCS module's portion of the end-to-end runtime exhibits the lowest memory usage, likely due to its efficient CSR graph representation. In contrast, peak memory consumption occurs during the degree sequence fitting stage in the case of RECCS+ on CEN clustered with Leiden with resolution 0.01, and during the SBM stages for all other configurations. Both high-memory scenarios involve graph-tool's general-purpose graph representation, which trades higher memory overhead for broader functionality and faster execution of graph operations. Future work could address overhead and memory consumption caused by Python process calls from C++, by reimplementing the Python processed in C++ and integrating them directly into the main RECCS++ runtime. This would eliminate the cost of spawning Python processes from C++ while enabling the use of more memory-efficient graph representations throughout the pipeline.

A more concrete limitation is that while we scaled up RECCSv1's explicit degree sequence fitting approach, we did not extend the same optimizations to RECCSv2, which uses SBMs for degree sequence fitting. Future work should investigate whether the parallelization strategies developed for RECCS+ can be effectively applied to SBM-based degree fitting, and whether the tradeoffs between accuracy and speed differ at scale. 

\section{Conclusion}\label{sec7}

In this paper, we have attained scalability through the use of shared memory parallelism and algorithmic reformulation of RECCS, resulting in RECCS+ and RECCS++. In networks such as OC and OCv2 where RECCS and RECCS+ could not complete, RECCS++ was able to complete within a twelve hour window, demonstrating its capability to handle previously intractable network scales.

%We have attained total speedups of 49× and 139× in RECCS+ and RECCS++ respectively, both in the case of CEN clustered with Leiden 0.01, where RECCS took 2 days, 5 hours and 24 minutes, RECCS+ took 1 hour and 4 minutes, and RECCS++ took 23 minutes.

We find that RECCS+ exhibits algorithmic fidelity with RECCS, and while RECCS++ exhibits reduced algorithmic fidelity, it maintains comparable network quality, showing several tradeoffs in metric accuracy measures. The reclustering accuracy confusion matrices reveal differences between the original RECCS networks and the RECCS+ and RECCS++ networks, particularly when IKC is used as the clustering or reclustering method. However, these differences reflect the fundamental tension between exact replication and practical scalability. The magnitude and nature of these differences also depend on the sensitivity of specific clustering algorithms to network perturbations—some methods like IKC may be more sensitive to minor structural variations than others.

Most importantly, RECCS+ and RECCS++ enable researchers to evaluate algorithms on network scales that were previously inaccessible, making it possible to assess algorithmic behavior in realistic, large-scale settings rather than being limited to smaller toy examples. Furthermore, the newfound scalability opens doors for new adaptations and applications of our synthetic network generator beyond the current scope of this work.

\backmatter

% \bmhead{Supplementary information}

\section*{Acknowledgements}

We thank the Insper-Illinois partnership for providing support for this project. We thank George Chacko, Tandy Warnow, The-Anh Vu Le, and Ian Wei Chen from the Siebel School of Computing and Data Science for helpful discussions and advice. This research was funded in part by NSF grant numbers CCF-2109988, OAC-2402560, and CCF-2453324.

\section*{Declarations}

Claude Sonnet 4.5, Claude Code, and GitHub Copilot, all AI assistants and agents created by Anthropic and GitHub, were used to assist in developing this software.

\bibliography{sn-bibliography}% common bib file

\end{document}